\input{aipcheck}

\documentclass[
    ,final            
  ]
  {aipproc}

\layoutstyle{6x9}

\begin{document}

\title{Baryon resonances as dynamically generated states in chiral dynamics}


\author{Daisuke Jido}{
  address={Yukawa Institute for Theoretical Physics, Kyoto University, Sakyo, 
  Kyoto 606-8502, Japan}
}

%

\begin{abstract}
We discuss baryon resonances which are dynamically generated in hadron 
dynamics based on chiral coupled channels approach.  With the dynamical description
of the baryon resonance, we discuss the origin of the resonance pole, finding
that for the description of $N(1535)$ some other components
than meson and baryon are necessary.  Since the chiral unitary model
provides a microscopic description in terms of constituent hadrons,
it is straightforward to calculate transition amplitudes and form factors
of resonances without introducing further parameters. Finally we briefly 
discuss few-body nuclear kaonic systems as hadronic molecular states.   
\end{abstract}

\maketitle


\section{Dynamical description of hadronic resonance}

In principle, all the hadron states are dynamically generated by quark and 
gluon fields obeying QCD.
Nevertheless, the current quarks and gluons appearing in QCD 
are not effective degrees of freedom 
for the understanding of the hadron structure. 
Thus, more efficient description of hadrons is favorable. There 
it is an important question what are the effective constituents in baryon resonances,
or what are active ingredients in dynamical description of baryon 
resonances.
Since baryon resonances are located where the strong decay channels are open, 
hadronic components are also important to understand the structure of 
the baryon resonances apart from the component originated 
from the quarks confined in the single potential. 
Therefore, for the investigation of the baryon resonance structure, the aspect of 
hadron dynamics should be unavoidably considered. 
%


In dynamical description, resonance states are obtained as solutions
of Schr\"odinger equation or Lippmann-Schwinger equation for scattering
matrix with a given Hamiltonian $H=H_{0}+V$, where $H_{0}$ is 
the Hamiltonian of free particles and $V$ represents the interaction of 
these particle. The Hamiltonian $H_{0}$ fixes the model space 
of dynamical elements, with which the resonance states are dynamically
described, and the potential $V$ specifies dynamics of the elementary
components. If states are obtained without introducing other components 
than those in $H_{0}$, we call these states dynamically generated state. 
If these ingredients are written in terms of hadrons, the resonance states 
are described by hadron dynamics. 

In the coupled channel approach with chiral dynamics (chiral unitary model),
the scattering amplitude is obtained as a solution of the Lippmann-Schwinger 
equation $T=V+VGT$ under the assumption that 
the model space spans the lowest lying octet mesons and baryons
and their interaction is given by chiral perturbation
theory~\cite{Kaiser:1995eg}. 
The off-shell behavior in the Lippmann-Schwinger equation is tamed with 
regularization of the integral by, for instance, introducing form factors
or using the dimensional regularization.
It is known that, based on the $N/D$ method, one can simplify the scattering 
equation and obtain an algebraic solution $T=(1-GV)^{-1}V$ by using the on-shell 
values of the potential $V$ and the loop function $G$ with 
the dimensional regularization~\cite{Oller:2000fj}.

Explicit pole terms in the interaction kernel $V$ represent states outside 
the model space. Thus, purely dynamically generated state should be obtained 
without the explicit pole terms. 
Even though the amplitudes are written in terms of hadron dynamics,
resonances generated dynamically are not always genuine hadronic composite objects. 
According to Ref.~\cite{Ecker:1988te}, in chiral perturbation theory, the low
energy constants in higher order terms are dominated by resonance contributions.
This means that the contributions of the resonances which are not 
described dynamically in the present model space are hidden in the interaction 
kernel and this kind of resonances can be reconstructed after dynamical 
calculation~\cite{Dobado:1996ps,Oller:1998hw}.
There is another source of contributions coming from outside the model space. 
In the regularization procedure, one fixes high-momentum behavior which is not 
controlled in the model space.
This means that some contributions coming from the outside of the model space 
can be hidden in the regularization parameters~\cite{Hyodo:2008xr}.

\section{Interpretation of resonance pole }

Here we discuss the interpretation of the resonance pole within the chiral unitary model. 
In the previous section, we have discussed the possible sources of resonances
from outside the model space. 
First of all, we discuss whether we exclude this source of the 
resonance from our description of the resonance theoretically. 
For the interaction kernel, since resonance contributions can be hidden
in the higher order terms of chiral perturbation theory, we take only the leading 
Tomozawa-Weinberg term, which is understood as the $t$-channel vector meson 
exchange. For the regularization parameter of the loop function $G$, it is possible 
to exclude the implicit source in a consistent way to the chiral 
expansion as the following way~\cite{Hyodo:2008xr}. 
If there are no states other than the free scattering
states in the loop function, the real part of the loop function should be negative
below the threshold, since the spectral representation of the Green function reads 
$ G(W) = \sum_{n} \rho(W)/(W-E_{n})$ with the positive definite spectral function 
$\rho(W)$ and the total energy $W$,
and the lowest states is the threshold at $W=E_{0}$, 
then ${\rm Re}G(W) \le 0$ for $W\le E_{0}$. In addition, if the chiral expansion 
is applied, at some point in the low-energy region the scattering amplitude can be written
in  chiral perturbation theory, namely $T=V$, which implies that $G=0$. 
Since the loop function is a decreasing function below the threshold,
these conditions can be satisfied at $W=M$ with the baryon mass $M$ by 
\begin{equation}
   G(M;a_{\rm natural}) = 0 , \label{eq:nat}
\end{equation}
which we call natural renormalization scheme~\cite{Hyodo:2008xr}. 
Equation (\ref{eq:nat}) fixes the renormalization constant in a consistent way 
with chiral expansion and exclusion of the resonance source. 

\begin{figure}
  \includegraphics[width=0.60\textwidth,bb=0 0 566 340]{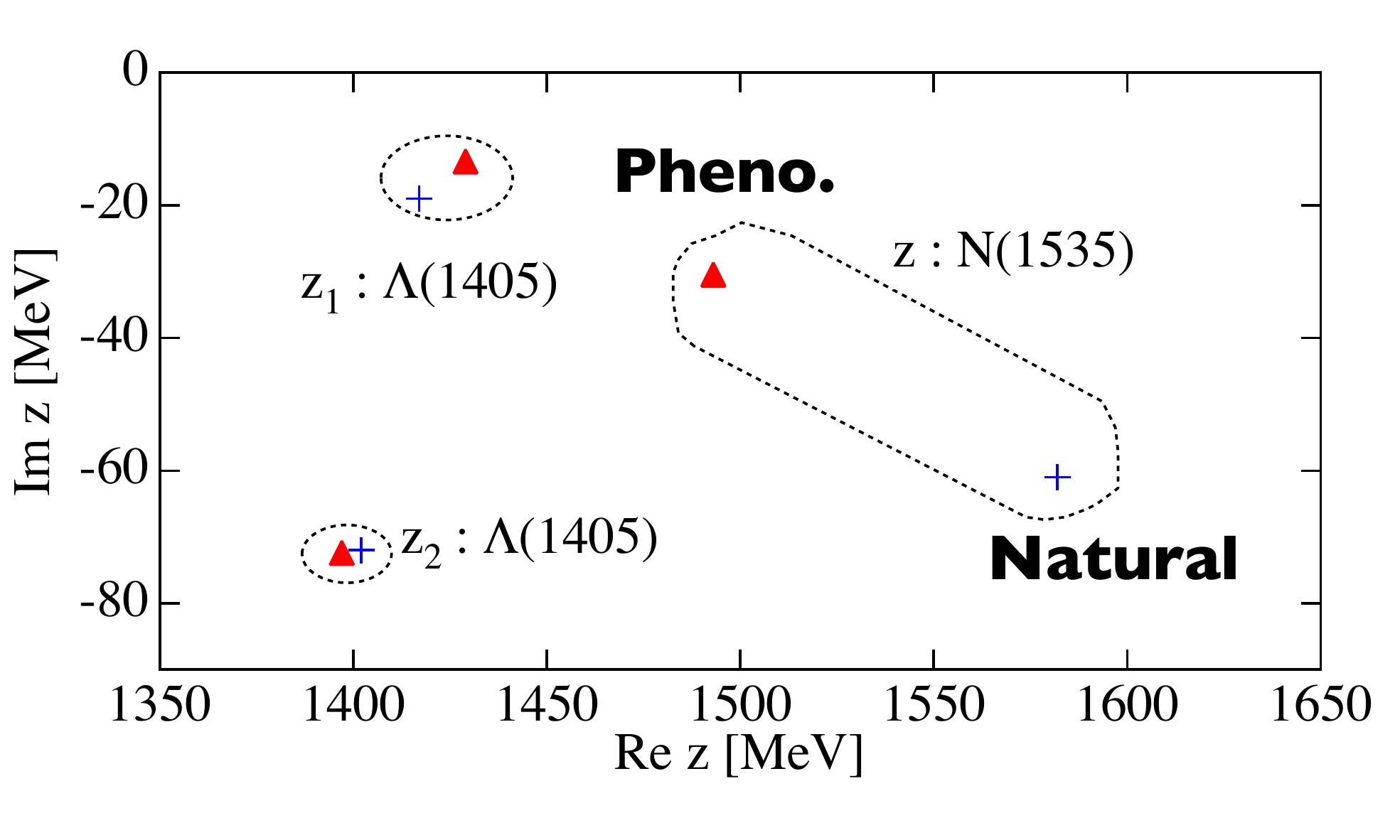}
  \caption{Comparison of the pole positions for $N(1535)$ and $\Lambda(1405)$
  obtained with the phenomenological and natural renormalization 
  schemes~\cite{Hyodo:2008xr}. $z$ denotes the pole positions of $N(1535)$ and 
  $z_{1}$ and $z_{2}$ are for the two states of $\Lambda(1405)$. The triangle 
  and cross mean the pole positions obtained by the phenomenological and natural
  renormalization schemes, respectively. }
  \label{fig:pole}
\end{figure}

If we solve the Lippmann-Schwinger equation with this renormalization 
parameter and the Tomozawa-Weinberg interaction $V_{TW}$, we obtain 
a dynamical description of the scattering amplitude in terms of hadrons as
\begin{equation}
T_{\rm natural}(W) = [V^{-1}_{TW}(W)-G(W;a_{\rm natural})]^{-1} , \label{natural}
\end{equation}
and poles appearing in this amplitude correspond to 
a genuine dynamically generated states in terms of hadron dynamics.
Nevertheless, it is not always the case that the scattering amplitude 
(\ref{natural}) and its poles agree with the experimental data. 
We discuss contribution from the outside of the model space  
by comparing the pole positions of the amplitude (\ref{natural})
and those of the scattering amplitude obtained phenomenologically so
as to reproduce the experimental data. The phenomenological amplitude
is obtained by using the Tomozawa-Weinberg interaction and 
the renormalization parameter determined by using experimental data: 
\begin{equation}
T_{\rm pheno}(W) = [V^{-1}_{TW}(W)-G(W;a_{\rm pheno})]^{-1} . \label{pheno}
\end{equation}
In Fig.~\ref{fig:pole}, we show the comparison of the pole 
positions for $N(1535)$ and $\Lambda(1405)$ obtained with the 
phenomenological and natural renormalization schemes. 
As one can see, for $N(1535)$, the two solutions differ from each other. 
This implies that, to describe the $N(1535)$ resonance, we need 
certain contributions coming from the components other than meson
and baryon, which are possibly quark-originated components. 
In contrast, for $\Lambda(1405)$, these two pole positions are
almost the same. This shows that the $\Lambda(1405)$ can be described 
dominantly by the meson-baryon component.

In the natural renormalization scheme with the Tomozawa-Weinberg interaction,
$\Lambda(1405)$ is successfully reproduced, while $N(1535)$ is obtained not 
so satisfactorily. 
It is interesting to see which kind of interaction is necessary
in the natural renormalization to reproduce a phenomenological description.
First we note that, in the renormalization point of view, once the 
scattering amplitude $T$ is fixed, the change of the renormalization parameter in 
$G$ should be absorbed into the interaction $V$. 
This implies that the equivalent scattering amplitude
can be expressed by different sets of $V$ and $G$ depending on the renormalization scheme labeled by $a$:
\begin{equation}
T(W) = [V^{-1}(W;a)-G(W;a)]^{-1}. \label{general}
\end{equation}
Now if we obtain a good phenomenological description of the scattering amplitude with an appropriate $a_{\rm pheno}$ as in Eq.~(\ref{pheno}) with the Tomozawa-Weinberg 
interaction,
we can obtain $V(W;a_{\rm natural})$ by equating Eqs.~(\ref{pheno}) and (\ref{general}). After some algebra, 
we find in a single channel case\footnote{Generalization to the coupled channel is straightforward and discussed in Ref.~\cite{Hyodo:2008xr}} 
\begin{equation}
V(W;a_{\rm natural})= V_{WT}(W) + \frac{C}{2f^2}\frac{(W-M)^2}{W-M_{\rm eff}}
\label{Vnatural}
\end{equation}
with $V_{TW}(W) = -C (W-M) /(2f^2)$, $M_{\rm eff} \equiv M - {2f^2}/({C \Delta a})$ and $\Delta a \equiv  G(W;a_{\rm natural})-G(W;a_{\rm pheno}) = a_{\rm natural}-a_{\rm pheno}$. 
As seen in Eq.~(\ref{Vnatural}),  the interaction kernel in the natural renormalization scheme is expressed by the WT term and a pole term with the mass $M_{\rm eff}$ depending on the difference of two renormalization schemes $\Delta a$.

The relevance of the pole term depends on the value of $M_{\rm eff}$. Using 
the values of $a_{\rm pheno}$
obtained in the coupled channel \cite{Hyodo:2002pk,Hyodo:2003qa} and 
taking a natural renormalization scheme $G(M_{N};a_{\rm natural}) = 0$ for all channels, 
we find the effective mass $M_{\rm eff} \approx 1700\pm 40i$ MeV for $N(1535)$\footnote{This value quantitatively has moderate dependence on the values of $a_{\rm pheno}$ and choice of the natural renormalization condition in the coupled channel.}.
Since this pole appears in the relevant energy of the $N(1535)$ physics,
it can be a source of the $N(1535)$ having some components other than 
dynamically generated state by meson and baryon. This pole
may be interpreted as a genuine quark component and could be a chiral partner 
of nucleon discussed in Refs.~\cite{Jido:1998av,Jido:2001nt}

\section{Applications of dynamical description}

Once we obtain a good and reliable description of hadron resonances, we can calculate
further their dynamical properties. So far, there have been, for instance,  
the investigations of 
the magnetic moments of $\Lambda(1405)$ and $\Lambda(1670)$~\cite{Jido:2002yz},
the radiative decay of $\Lambda(1405)$~\cite{Geng:2007hz},
the helicity amplitudes of $N(1535)$, $\Lambda(1405)$ and $\Lambda(1670)$~\cite{Jido:2007sm,Doring:2010rd}
and 
the electromagnetic form factors of $\Lambda(1405)$~\cite{Sekihara:2008qk,Sekihara:2010uz}.
There are many works for reaction calculations with the dynamical description 
based on chiral dynamics.

Since the resonance is described dynamically in terms of the constituent meson
and baryon microscopically, it is straightforward to calculate the helicity amplitude
and form factors of the resonance by implementing
the photon coupling to the constituent hadrons. Once we fix the elementary 
couplings of the meson and hadrons to the photon, there are no additional 
parameters.
The helicity amplitudes of $N(1535)$ have been discussed 
in the chiral unitary model~\cite{Jido:2007sm}, in which the $N(1535)$ is 
described with the phenomenological renormalization scheme, and 
the electromagnetic transitions $\gamma^{*}N \rightarrow N(1535)$ are calculated.  
It is very interesting that this model reproduces the observed helicity amplitudes, 
especially the neutron-proton ratio $A_{1/2}^{n}/A_{1/2}^{p}$ in good agreement
with experiment, although this model implicitly has the quark-originated  
pole for $N(1535)$ as discussed above and the direct photon couplings to the
quark components were not considered in this calculation. 
Therefore, the success of this model implies that the
meson-baryon components of $N(1535)$ are essential
for the structure of $N(1535)$ probed by a low-energy virtual photon. 
The effect of the quark core and its relation to the meson cloud are discussed in 
Refs.~\cite{Ramalho:2011ae,Ramalho:2011fa}.

The electromagnetic form factors of $\Lambda(1405)$ were also calculated 
within the chiral unitary model using the chiral effective theory for the 
couplings of the external currents with the hadronic 
constituents~\cite{Jido:2002yz,Sekihara:2008qk,Sekihara:2010uz}.
The first moment of the form factor, which corresponds to the 
mean-squared radius in the limit that the resonance width goes to zero,
was also calculated. 
The electric first moment of $\Lambda(1405)$ at the pole position was 
obtained as a complex number $-0.13 + 0.30i$ fm$^{2}$, 
whose modules, 0.33 fm$^{2}$, is much larger than the neutron charge radius. 
This may imply that the $\Lambda(1405)$ has a spatially larger size than
the typical hadronic size.


The dynamical description of  $\Lambda(1405)$ 
shows that one of the $\Lambda(1405)$ states 
is almost a bound state of $\bar KN$. The scalar mesons, $f_{0}(980)$
and $a_{0}(980)$, have been also considered to have large components
of $\bar KK$.   A state which is essentially 
described by hadronic components is called hadronic molecular state.
In the hadronic molecular states, constituent hadrons keep their identities 
as they are in isolated systems. 
Estimating 
the strength of the inter-hadron interactions between $\bar KN$ and $\bar KK$ 
by chiral effective theory, in which the Weinberg-Tomozawa
interaction is responsible for the low-energy $s$-wave $\bar KN$ and $\bar KK$
attractions,
we see that this interaction is strong enough to produce $\bar KN$ and $\bar KK$
bound states with a few tens MeV binding energies. 
If one compares this chiral 
effective interaction with the $NN$ interaction, the $\bar KN$ and $\bar KK$ 
attractions are very strong, because the $NN$ bound state, that is deuteron, 
has as small as 2 MeV binding energy. But if one compares 
the binding energies to the typical hadronic scale of several hundred MeV, one 
should say that these inter-hadron interactions are weak. 

It is also interesting to mention the fact that 
the $\bar KN$ and $\bar KK$ attractions obtained from the 
Tomozawa-Weinberg interaction have very similar strengths,
because the strength of the 
Tomozawa-Weinberg interaction is given by the SU(3) flavor symmetry and 
$K$ and $N$ are classified into the same state vector in the octet representation. 
This similarity between $K$ and $N$ leads to 
systematics of three-body kaonic systems~\cite{Hyodo:2011ur}, 
$\bar KNN$, $\bar KKN$ and $\bar KKK$.
%
%
The $\bar KKN$ quasibound state is an important resonance for the $N^{*}$ 
physics, since it has the same quantum number as $N^{*}(P_{11})$. 
This state was studied first with a simple  
single-channel non-relativistic potential model~\cite{Jido:2008kp} and
later  was investigated in a more 
sophisticate calculation~\cite{MartinezTorres:2008kh,MartinezTorres:2010zv}
based on a coupled-channels Faddeev method and
a simple fixed center approximation
of three-body calculation~\cite{Xie:2010ig}.  
These approaches lead to a very similar $N^{*}$ resonance state appearing
around 1910 MeV. 
%
The potential model calculation shows that 
the root mean-squared radius of the $\bar KNN$ state is as large as 1.7 fm, 
which are similar with the radius of $^{4}$He.
The inter-hadron distances are comparable with an average 
nucleon-nucleon distance in nuclei. Thus, this $N^{*}$ resonance has an
much larger spatial size than typical $N^{*}$ resonances which are made
of constituent quarks confined in 1 fm. 
The hadronic molecular states could be identified 
by production rates in heavy ion collisions,
since coalescence of hadrons to produce 
loosely bound hadronic molecular systems is more probable 
than quark coalescence for compact multi-quark systems~\cite{Cho:2010db}.

%
%
%
%

%

\section{Summary}

Coupled-channel approaches, for instance the chiral unitary model, 
provide us with a dynamical description of meson-baryon scattering,
describing simultaneously both resonance and nonresonant scattering  
applicable to reaction calculations.
They also give us hadronic description in which 
all contents of the models are hadrons. 
Nevertheless, obtained hadron resonances are not necessarily genuine hadronic 
composite objects and sources of quark dynamics can be hidden everywhere. 
Thus, detailed theoretical analyses are necessary to interpret the structure of resonances. 
In the chiral unitary model, the resonance can be described 
microscopically in terms of the constituent hadrons based on  
fundamental interactions given in the chiral effective theory.
This is well suited to the calculation of form factors of resonances. 
For the effective constituents in baryons structure
hadrons themselves can be effective constituents in some hadron resonances,
for instance few-body systems with nucleons and kaons.






\begin{theacknowledgments}
The author would like to thank his collaborators for their collaborations on 
the works presented here.   
This work was partially supported by the Grant-in-Aid for Scientific Research from 
MEXT and JSPS (Nos.\ 22740161, 22105507).
This work was done in part under the Yukawa International Program for Quark-hadron Sciences (YIPQS).
\end{theacknowledgments}





\IfFileExists{\jobname.bbl}{}
 {\typeout{}
  \typeout{******************************************}
  \typeout{** Please run "bibtex \jobname" to optain}
  \typeout{** the bibliography and then re-run LaTeX}
  \typeout{** twice to fix the references!}
  \typeout{******************************************}
  \typeout{}
 }

\end{document}